\documentclass[aps,prb,superscriptaddress,showpacs,twocolumn]{revtex4}
\usepackage{amsfonts}
\usepackage{amsmath}
\usepackage{amssymb}
\usepackage{graphicx}

\begin{document}

\title{Spin transfer torques in nonlocal lateral spin valve}
\author{Yuan Xu}
\affiliation{Institute of Physics, Chinese Academy of Sciences, Beijing 100080, China}
\author{Ke Xia}
\affiliation{Institute of Physics, Chinese Academy of Sciences, Beijing 100080, China}
\author{Zhongshui Ma}
\affiliation{School of Physics, Peking University, Beijing 100871, China}
\date{\today }

\begin{abstract}
We report a theoretical study on the spin and electron transport in
the nonlocal lateral spin valve with non-collinear magnetic
configuration. The nonlocal magnetoresistance, defined as the
voltage difference on the detection lead over the injected current,
is derived analytically. The spin transfer torques on the detection
lead are calculated. It is found that spin transfer torques are
symmetrical for parallel and antiparallel magnetic configurations,
which is different from that in conventional sandwiched spin valve.
\end{abstract}

\maketitle

\section{Introduction}

Because of increasing interests in nano-structures with a spin
degree of freedom incorporated, the local spin valve (LSV), where
a layer of normal metal (NM) or insulator is sandwiched by two
layers of ferromagnetic metal (FM), has been considered as the
prototype of experimental setup for demonstration
of spin dependent effects, such as GMR,\cite{GMR} magnetization switching,%
\cite{switch_myers,Slonczewski,Berger} \textit{etc}. However, it
is not easy for precisely analyzing the spin transport based on
LSV in experiment. The reason is that, accompanying the electrical
current flowing across the detection ferromagnetic contact, the
spurious effects such as anisotropy magnetoresistance and Hall
effect due to the FM contact are also
involved.\cite{Jedema_nature_2001} This problem can partially
removed by using nonlocal lateral spin valve (NLSV), where only
spin current flows across the detection FM
contact.\cite{Jedema_nature_2001} Recently, several experiments of
metallic spin injection and detection had been carried out on
NLSV.\cite{Mark_Johns_PRL_1985,Mark_Johns_PRB1_1985,
Johnson_PRL_2006,
Joonyeon_ARL_2006,Jedema_nature_2001,Jedema_nature_2002,
Jedema_PRB_2003,Ji_NLSV} From these experiments, important
parameters
of spin transport, such as spin diffusion length, are obtained.\cite%
{Diffusion_review_J_Bass}

However, most of those experiments focused on collinear magnetic
configuration, in which the magnetization of injection source and
detection drain are arranged to be parallel or antiparallel. On
the other hand, the noncollinear spin
transport in LSV has been studied extensively\cite%
{Urazhdin_AMR,Universial_AMR,Noncollinear_Hernando,Kovalev_STT,Manschot_STT,Waintal_STT}
and reveals interesting physics, such as spin transfer torques (STT) and
related magnetization switching. Little effort has been put on the
noncollinear spin transport in NLSV so far. For NLSV, it is interesting to
know whether or not we can also obtain \emph{sizable} STT, and how the spin
current behaves when carried by the diffusion of spin instead of the
electrically assistant drift of spin. Recently, the current induced
magnetization switching was realized in NLSV,\cite{Kimura_NLSV_switch} which
gave a strong evidence of the presence of STT effect even in NLSV.

In this paper, by combining the diffusion equation and the
magnetoelectronic circuit
theory,\cite{Brataas_circuit,Circuit_theory_review} we investigate
theoretically the spin transport in NLSV with non-collinear
magnetic configurations. The angular magnetoresistance (AMR) in
NLSV is discussed for systems with metallic FM$/$NM contacts and
tunnelling contacts. It is also shown that because of the spin
accumulation at the normal metal side of FM$/$NM contact, STT
could be acted on the ferromagnet. When the length of NM stripe is
less than the spin diffusion length in NM, we found that STT in
NLSV is comparable with that in LSV. The angular dependence of the
torques is qualitatively different from that in LSV.

The paper is organized as follows. In Sec.II, the theoretical
frame for dealing with the non-collinear transport in NLSV is
presented and analytical expressions for both AMR and STT are
derived. In Sec.III, we calculate the AMR and STT in the NLSV, the
properties of torques and voltage difference across the FM$/$NM
junction are discussed. Finally, we summarized our paper in Sec.
IV.

\section{Theory description and models}

\begin{figure}[tbp]
\includegraphics[width=7cm]{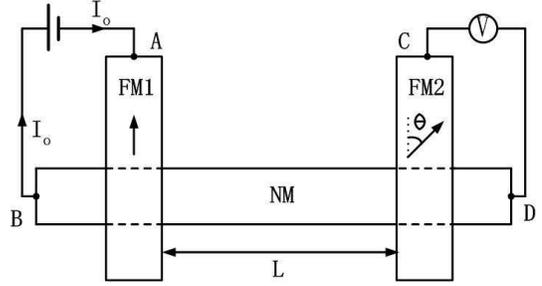}
\caption{Experimental set up of NLSV, where $L$ denotes the edge
to edge space, $\protect\theta$ is the relative angle between the
magnetization of the two FM leads and A,B,C,D denote the
electrodes connecting outer circuit. $I_{o}$ gives the electron
(particle) current.} \label{fig:1}
\end{figure}

Fig.\ref{fig:1} is the schematic of the NLSV experiment setup. It
consists of one NM lead and two ferromagnetic leads FM1 and FM2.
These two ferromagnetic leads are separated by a length $L$ and
are aligned parallel to each other. Experimentally, the current
$I_{o}$ is injected from FM1 and flows out from the left end of
NM. In this work, the direction of electrical current $I_{o}$ is
defined along the direction of the electron (particle) current.
After the injection, the spin is accumulated in the NM lead. The
diffusive spin spreads over the region in the NM lead between the
two FM$/$NM contacts. A voltage difference $V$ across FM2$/$NM
contact could be built up.\cite {Mark_Johns_PRL_1985,
Mark_Johns_PRB1_1985} For the different configurations of
magnetization arrangements, \textit{i.e.}, with different $\theta
$ in Fig.\ref{fig:1}, the spin-accumulation induced voltage across
FM2$/$NM contact is angle-dependent($V(\theta )$). It can be
measured by the nonlocal AMR defined as $R(\theta )\equiv V(\theta
)/I_{o}$.

\subsection{Theoretical frame of transport in NLSV}

The transport theory in a NLSV should include three parts, the
transport in FM, NM resistors and across the FM$/$NM contacts. As
the dimensions of FM and NM resistors in NLSV typically are much
larger than electron mean free path, the transport can be
described by the diffusion equation in terms of the spatially
dependent electrochemical potential.\cite{Datta95}

In spin polarized system, besides the electrochemical potential $%
u_{0}\left( x\right) =u_{ch}(x)-e\phi (x)$ where $u_{ch}$ gives
the chemical potential and $\phi $ gives the electric
potential($-e$ denoting the electron charge), it is necessary to
introduce a quantity $\mathbf{u}_{s}(x)$
accounting for the spin accumulation in the system.\cite%
{Brataas_circuit,Circuit_theory_review} The direction of
$\mathbf{u}_{s}(x)$ denotes the direction of spin accumulation in
spin space and the magnitude of $\mathbf{u}_{s}(x)$ gives the
energy splitting of the two spins in local coordinate system. In
principle, the direction of $\mathbf{u}_{s}(x)$ in the normal
metal is arbitrary and need to be determined by boundary
conditions. In a ferromagnet, the spin accumulation reads
$\mathbf{u}_{s}^{F}\left( x\right) =\mathbf{m}(u_{\uparrow
}^{F}\left( x\right) -u_{\downarrow }^{F}\left( x\right) )$, where
$\mathbf{m}$ is an unit vector along the magnetization in FM and
$u_{\uparrow \left( \downarrow \right) }^{F}\left( x\right) $ is
the electrochemical potential of majority (minority) spin in the
local coordinate system where the quantized axis is parallel to
the magnetization.

\textbf{\emph{Transport in FM}}: As the spin decoherence length is
of the order of lattice constants in conventional
ferromagnet,\cite{spin_MD_Stile} only the components which are
parallel or antiparallel to the magnetization direction
$\mathbf{m}$ in FM can survive. Therefore, the electrical and the
spin currents in FM region
are\cite{Hershfield,Takahashi_spin_injection} $I_{0}^{F}\left(
x\right) =-(S^{F}/e)(\sigma _{\uparrow }^{F}\nabla _{x}u_{\uparrow
}^{F}\left( x\right) +\sigma
_{\downarrow }^{F}\nabla _{x}u_{\downarrow }^{F}\left( x\right) )$ and $%
\mathbf{I}_{s}^{F}\left( x\right) =-(S^{F}/e)\nabla _{x}(\sigma _{\uparrow
}^{F}u_{\uparrow }^{F}\left( x\right) -\sigma _{\downarrow
}^{F}u_{\downarrow }^{F}\left( x\right) )\mathbf{m}$. Here the transport is
assumed to be along $x$ axis, $S^{F}$ is the area of the cross section in FM
and $\sigma _{\uparrow (\downarrow )}^{F}$ denotes the conductivity for
majority (minority) spin channel. The bulk parameters, such as conductivity $%
\sigma $, are assumed to be spatially uniform in this study.

Correspondingly, with conservation of electrical current $I_{0}^{F}$, the
continuity equations of the spin current are\cite{Hershfield} $\nabla
_{x}I_{\uparrow }^{F}\left( x\right) /S^{F}=-e\xi _{\uparrow }(u_{\uparrow
}^{F}\left( x\right) -u_{0}^{F}\left( x\right) )/\tau _{\uparrow \downarrow
}+e\xi _{\downarrow }(u_{\downarrow }^{F}\left( x\right) -u_{0}^{F}\left(
x\right) )/\tau _{\downarrow \uparrow }$ and $\nabla _{x}I_{\downarrow
}^{F}\left( x\right) /S^{F}=-e\xi _{\downarrow }(u_{\downarrow }^{F}\left(
x\right) -u_{0}^{F}\left( x\right) )/\tau _{\downarrow \uparrow }+e\xi
_{\uparrow }(u_{\uparrow }^{F}\left( x\right) -u_{0}^{F}\left( x\right)
)/\tau _{\uparrow \downarrow }$, where $\xi _{\uparrow (\downarrow )}$ is
the density of states per unit volume at Fermi level for single spin, $\tau
_{\uparrow \downarrow }$ and $\tau _{\downarrow \uparrow }$ are the
spin-flip scattering time for majority and minority spins.

Inserting the expression of spin current into the continuity
equations and with detailed balance $\xi _{\uparrow }/\tau
_{\uparrow \downarrow }=\xi _{\downarrow }/\tau _{\downarrow
\uparrow }$, we obtain the
conjugated diffusion equations for $u_{\uparrow }^{F}\left( x\right) $ and $%
u_{\downarrow }^{F}\left( x\right) $ in ferromagnetic metal as
\begin{equation}
\nabla _{x}^{2}u_{\uparrow }^{F}\left( x\right) =u_{\uparrow }^{F}\left(
x\right) /D_{\uparrow }\tau _{\uparrow \downarrow }-u_{\downarrow
}^{F}\left( x\right) /D_{\uparrow }\tau _{\uparrow \downarrow },
\end{equation}%
\begin{equation}
\nabla _{x}^{2}u_{\downarrow }^{F}\left( x\right) =-u_{\uparrow }^{F}\left(
x\right) /D_{\downarrow }\tau _{\downarrow \uparrow }+u_{\downarrow
}^{F}\left( x\right) /D_{\downarrow }\tau _{\downarrow \uparrow },
\end{equation}%
where $D_{\uparrow (\downarrow )}$ is the diffusion constant for
majority (minority) spin and relates to $\sigma _{\uparrow
(\downarrow )}^{F}$ via the Einstein relation $\sigma _{\uparrow
(\downarrow )}^{F}=e^{2}\xi _{\uparrow (\downarrow )}D_{\uparrow
(\downarrow )}$.\cite{Datta95} Solving the diffusion
equations, we obtain the spin-resolved electrochemical potential in FM\cite%
{Hershfield}
\begin{eqnarray}
\left(
\begin{array}{c}
u_{\uparrow }(x) \\
u_{\downarrow }(x)%
\end{array}%
\right)  &=&(\widetilde{A}+\widetilde{B}x)\left(
\begin{array}{c}
1 \\
1%
\end{array}%
\right) +\widetilde{C}e^{x/l_{sf}^{F}}\left(
\begin{array}{c}
\sigma _{\uparrow }^{F-1} \\
-\sigma _{\downarrow }^{F-1}%
\end{array}%
\right)   \notag \\
&&+\widetilde{D}e^{-x/l_{sf}^{F}}\left(
\begin{array}{c}
\sigma _{\uparrow }^{F-1} \\
-\sigma _{\downarrow }^{F-1}%
\end{array}%
\right) ,  \label{eq-21}
\end{eqnarray}%
where $\widetilde{A}$, $\widetilde{B}$, $\widetilde{C}$, $\widetilde{D}$ are
constants to be determined by boundary conditions, and $l_{sf}^{F}$ is the
spin diffusion length in FM given by $[(D_{\uparrow }\tau _{\uparrow
\downarrow })^{-1}+(D_{\downarrow }\tau _{\downarrow \uparrow })^{-1}]^{-1/2}
$.

\textbf{\emph{Transport in NM}}: The electrical and spin currents in NM are
also governed by the diffusion equation as\cite{Brataas_circuit} $%
I_{0}^{N}\left( x\right) =-(\sigma ^{N}/e)S^{N}\nabla
_{x}u_{0}^{N}\left( x\right) $ and $\mathbf{I}_{s}^{N}\left(
x\right) =-(\sigma ^{N}/2e)S^{N}\nabla
_{x}\mathbf{u}_{s}^{N}\left( x\right) $. $S^{N}$ is the cross
section of NM and $\sigma ^{N}$ is the conductivity of NM.
Conservation of electrical current requires $\nabla
_{x}I_{0}^{N}\left( x\right) =0$ which leads to $\nabla
_{x}^{2}u_{0}^{N}\left( x\right) =0.$ Experimentally, the sample
length of NM is always comparable or longer than spin diffusion
length in NM. Therefore, the spin-flip scattering can not be
neglected. The continuity condition of spin current in NM reads
$\left( 1/S^{N}\right) \nabla _{x}\mathbf{I}_{s}^{N}\left(
x\right) =-e\left( \xi ^{N}/2\right) \mathbf{u}_{s}^{N}\left(
x\right) /\tau _{sf}^{N} $, where $\xi ^{N}$ is the total density
of states per unit volume at Fermi level in NM and $\tau
_{sf}^{N}$ is the spin relaxation time in NM. With the
current and continuity equation, the diffusion equation for $\mathbf{u}%
_{s}^{N}\left( x\right) $ reads
\begin{equation}
\nabla _{x}^{2}\mathbf{u}_{s}^{N}\left( x\right) =\mathbf{u}_{s}^{N}\left(
x\right) /(l_{sf}^{N})^{2},
\end{equation}%
where $l_{sf}^{N}=(D^{N}\tau _{sf}^{N})^{1/2}$ is the spin diffusion length
in NM. $D^{N}$ is diffusion constant and related to $\sigma ^{N}$ via $%
\sigma ^{N}=e^{2}\xi ^{N}D^{N}$. Solving the diffusion equation,
the spin accumulation in the NM can be written in the form as
\begin{equation}
\mathbf{u}_{s}^{N}\left( x\right) =\widetilde{\mathbf{E}}e^{x/l_{sf}^{N}}+%
\widetilde{\mathbf{F}}e^{-x/l_{sf}^{N}}
\end{equation}%
where $\widetilde{\mathbf{E}}$ and $\widetilde{\mathbf{F}}$ are the constant
vectors depending on the boundary conditions.

\textbf{\emph{Transport across FM$/$NM}}: In the absence of the interfacial
spin-flip scattering, the electrical current $I_{0}^{N|F}$ and the spin
current $\mathbf{I}_{s}^{N|F}$ across the FM$/$NM contact, which are
evaluated at the NM side, can be written in terms of electrochemical
potential and spin accumulation in linear response regime as\cite%
{Brataas_circuit}

\begin{eqnarray}
eI_{0}^{N|F} &=&(G_{\uparrow }^{I}+G_{\downarrow }^{I})(u_{0}^{N}\left(
x_{I}^{-}\right) -u_{0}^{F}\left( x_{I}^{+}\right) )  \notag \\
&&+\frac{1}{2}(G_{\uparrow }^{I}-G_{\downarrow }^{I})(\mathbf{m}\cdot
\mathbf{u}_{s}^{N}\left( x_{I}^{-}\right) -u_{s}^{F}\left( x_{I}^{+}\right) )
\end{eqnarray}%
and%
\begin{eqnarray}
e\mathbf{I}_{s}^{N|F} &=&\mathbf{m}[(G_{\uparrow }^{I}-G_{\downarrow
}^{I})(u_{0}^{N}\left( x_{I}^{-}\right) -u_{0}^{F}\left( x_{I}^{+}\right) )
\notag \\
&&-\frac{1}{2}(G_{\uparrow }^{I}+G_{\downarrow }^{I})u_{s}^{F}\left(
x_{I}^{+}\right)   \notag \\
&&-\frac{1}{2}(2\text{Re}G_{\uparrow \downarrow }^{I}-G_{\uparrow
}^{I}-G_{\downarrow }^{I})\mathbf{m}\cdot \mathbf{u}_{s}^{N}\left(
x_{I}^{-}\right) ]  \notag \\
&&+\text{Re}G_{\uparrow \downarrow }^{I}\mathbf{u}_{s}^{N}\left(
x_{I}^{-}\right) -\text{Im}G_{\uparrow \downarrow }^{I}\mathbf{m}\times
\mathbf{u}_{s}^{N}\left( x_{I}^{-}\right) ,  \notag \\
&&  \label{eq-10}
\end{eqnarray}%
where $u_{0}^{F}\left( x_{I}^{+}\right) =(u_{\uparrow }^{F}\left(
x_{I}^{+}\right) +u_{\downarrow }^{F}\left( x_{I}^{+}\right) )/2$, the index
'$I$' refers to the contact, $x_{I}^{+(-)}$ denotes the position in the
immediate vicinity of the contact at the FM(NM) side. $G_{\uparrow
(\downarrow )}^{I}$ is the conductance of FM$/$NM contact for the majority
(minority) spin, and the complex quantity $G_{\uparrow \downarrow }^{I}$ is
the mixing conductance describing the non-collinear transport.\cite%
{Brataas_circuit} In a metallic system, the imaginary part of $G_{\uparrow
\downarrow }^{I}$ is usually two orders less than the real part\cite%
{torque_Ke} and will be neglected in this work.

\textbf{\emph{Boundary conditions}}: In the steady state, the charge
accumulation across the FM$/$NM contact is invariant, which leads to the
conservation of electrical current across the contact as
\begin{equation}
I_{0}^{N}\left( x_{I}^{-}\right) =I_{0}^{N|F}=I_{0}^{F}\left(
x_{I}^{+}\right) .
\end{equation}%
The transverse spins injected into FM are suppressed in the scale of spin
decoherence length\cite{spin_MD_Stile} and the component of spin
accumulation collinear with magnetization direction $\mathbf{m}$ of FM
should keep invariant in the steady state, which gives the conservation of
spin current collinear with magnetization across the contact as
\begin{equation}
\mathbf{m}\left( \mathbf{m}\cdot \mathbf{I}_{s}^{N}\left( x_{I}^{-}\right)
\right) =\mathbf{m}\left( \mathbf{m}\cdot \mathbf{I}_{s}^{N|F}\right) =%
\mathbf{I}_{s}^{F}\left( x_{I}^{+}\right) .
\end{equation}

In the adiabatic approximation, the suppression of non-collinear part of
spin current, in turn, results in the angular momentum to be transferred
into the local magnetic moment in FM. As the consequence, the STT on the
ferromagnet generated by the spin current can be expressed as

\begin{equation}
\tau =-\frac{\hbar }{2e}[\mathbf{I}_{s}^{N|F}-\mathbf{m}(\mathbf{m}\cdot
\mathbf{I}_{s}^{N|F})].  \label{eq-3}
\end{equation}%
STT could raise an additional term in the Landau-Lifshitz-Gilbert equation
as $\left. \partial _{t}\mathbf{m}\right\vert _{STT}=-\frac{\gamma }{M_{s}V}%
\tau $, where $\gamma >0$ is the gyromagnetic ratio\ and $M_{s}$ is the
magnitzaton and $V$ is the volume of the ferromagnet.

\subsection{The nonlocal AMR and STT}
To consider the nonlocal AMR defined as $V(\theta)/I_{o}$, with the
current $I_{o}$ in FM1 as input condition we need to know the
voltage over FM2/NM contact. By solving the diffusion equations with
the boundary conditions, the spatial distribution of electrochemical
potentials in FM and NM resistors can be obtained. For FM2 lead, the
local electrochemical potential far from the FM2/NM contact
($x\rightarrow\infty$) gives the experimentally measured voltage
across the FM2/NM as $V=u^{F}(\infty)$/(-e), where the zero
potential set at the NM side of FM2/NM. Then, the angular dependence
of the nonlocal AMR can be obtained analytically as
\begin{widetext}
\begin{equation}\label{eq-6}
R({\theta})=\frac{2R_{N}e^{-L/l_{sf}^{N}}cos\theta\prod^{2}_{i=1}
(P^{I}_{i}\eta^{I}_{i}+\alpha^{F}_{i}\eta^{F}_{i})}{e^{-2L/l_{sf}^{N}}-\prod^{2}_{i=1}
(2\eta^{I}_{i}+2\eta^{F}_{i}+1)+sin^{2}\theta[1-e^{2L/l_{sf}^{N}}\prod^{2}_{i=1}(2\rho^{I}_{i}+1)]
^{-1}\prod^{2}_{i=1}(2\eta^{I}_{i}+2\eta^{F}_{i}-2\rho^{I}_{i}) }
,
\end{equation}
\end{widetext}where the subindex $i=1(2)$ denotes ferromagnetic
lead FM1(FM2) and the corresponding contact FM1$/$NM (FM2$/$NM).
We have introduced
three dimensionless quantities, $\eta ^{I}=R^{I}/[(1-(P^{I})^{2})R^{N}]$, $%
\eta ^{F}=R^{F}/[(1-(\alpha _{F})^{2})R^{N}]$, and $\rho
^{I}=(2\text{Re}G_{\uparrow \downarrow }^{I})^{-1}/R^{N}$, with
interfacial resistance $R^{I}\equiv (G_{\uparrow }^{I}+
G_{\downarrow }^{I})^{-1}$, $R^{F}\equiv l_{sf}^{F}/(\sigma
^{F}S^{F})$ and $R^{N}\equiv l_{sf}^{N}/(\sigma ^{N}S^{N}) $ are the
resistances in FM and NM within the range of non-equilibrium spin
accumulation relaxations length. $P^{I}=(G_{\uparrow
}^{I}-G_{\downarrow }^{I})/(G_{\uparrow }^{I}+G_{\downarrow }^{I})$
is the polarization across the contact. $\sigma ^{F}=\sigma
_{\uparrow }^{F}+\sigma _{\downarrow }^{F}$ and $\alpha ^{F}=(\sigma
_{\uparrow }^{F}-\sigma _{\downarrow }^{F})/(\sigma _{\uparrow
}^{F}+\sigma _{\downarrow }^{F})$ are the conductivity and
polarization in the ferromagnet, respectively. For the cases of
$\theta =0$
or $\theta =180^{o}$, Eq.(\ref{eq-6}) reduces to previous result\cite%
{Takahashi_spin_injection} exactly.

The angular dependence of $R(\theta)$ is introduced by the cosine
function on the numerator and the term containing $sin^{2}\theta$ on
the denominator. As will be illustrated in the next part, the cosine
function gives the configuration symmetry between the two leads
while the $sin^{2}\theta$ related term describes the noncollinear
transport across the FM/NM contact. If FM/NM contact does not
dominate the transport of the circuit, $sin^{2}\theta$ related term
will not give obvious effect on the angular dependence and
$R(\theta)$ takes the form of cosine function.

According to Eq.(\ref{eq-6}), the increase of interfacial
polarization $P^{I}$ and ferromagnetic polarization $\alpha_{F}$
could increase AMR, as in this case the injected spin accumulation
in NM resistor could be enhanced. In the limit of heavy spin-flip
scattering in NM resistor, namely $l_{sf}^{N}\rightarrow0$,
Eq.(\ref{eq-6}) gives the vanishing AMR, which is expected as the
spin accumulation is completely consumed in NM resistor.

The analytical result obtained in Eq.(\ref{eq-6}) is universal for
the diffusive metallic systems without spin-flip scattering at
contacts. For a special case with tunnelling contacts (e.g., with
several oxidant metallic layers located at
contact\cite{Jedema_nature_2002}), the transport
properties of system are dominated by the contact as $R^{I}>>R^{N}(R^{F})$%
,which means $\eta ^{I}>>\eta ^{F}$ in our formulism. Then, AMR for
tunnelling contact is found to be

\begin{equation}
R({\theta })=-\frac{1}{2}\frac{P_{1}^{I}P_{2}^{I}R_{N}e^{-L/l_{sf}^{N}}cos%
\theta }{1-sin^{2}\theta \lbrack 1-e^{2L/l_{sf}^{N}}\prod_{i=1}^{2}(2\rho
_{i}^{I}+1)]^{-1}}.  \label{eq-7}
\end{equation}

For any type of contact, following Eq.(\ref{eq-3}), the STT exerted on FM2
is obtained formally as
\begin{equation}
\tau =-\frac{\hbar }{2e^{2}}\text{Re}G_{\uparrow \downarrow }^{I}\mathbf{m}%
_{2}\times \mathbf{u}_{s}^{N}(x_{I_{2}}^{-})\times \mathbf{m}_{2},
\label{eq-8}
\end{equation}%
where $\mathbf{m}_{2}$ denotes the direction of magnetization in
FM2, $\mathbf{u}_{s}^{N}(x_{I_{2}}^{-})$ is the spin accumulation
at the NM side of FM2$/$NM. According to Eq.(\ref{eq-8}), the STT
on FM2 is proportional to the spin accumulation, which restores
the form of STT in LSV.\cite{Circuit_theory_review} The magnitude
and the direction of spin accumulation in NM should be solved with
the help of the boundary conditions both at FM1$/$NM and FM2$/$NM
contacts.

STT $\tau $ given in Eq.(\ref{eq-8}) could be formally rewritten
as\cite{Slonczewski}
\begin{equation}
\tau =-\delta (\theta )I_{o}(\mathbf{m}_{2}\times
\mathbf{m}_{1}\times \mathbf{m}_{2}),  \label{eq-9}
\end{equation}%
where $I_{o}$ is the electron current and $\delta (\theta )$
yields an effective spin torques, which directly scales the
critical current of magnetization switching and switching time in
dynamics.\cite{Review_J_Z_Sun} The analytic expression for $\delta
(\theta )$ read as
\begin{equation}\label{eq-11}
\delta (\theta )= \mathcal{T}\frac{\hbar}{2e}\text{Re}G_{\uparrow
\downarrow }^{I}\frac{R(\theta)}{cos(\theta)}
\end{equation}
where the angular independent coefficient $
\mathcal{T}=2\rho_{2}^{I}\Phi/\Omega$, where

\begin{eqnarray}
\Phi=1+e^{4L/l_{sf}^{N}}\prod^{2}_{i=1}(2\rho_{i}^{I}+1)(2\eta_{i}^{I}+2\eta_{i}^{F}+1)\notag \\
-e^{2L/l_{sf}^{N}}\prod^{2}_{i\neq
j}(2\rho_{i}^{I}+1)(2\eta_{j}^{I}+2\eta_{j}^{F}+1)
\end{eqnarray}
and
\begin{eqnarray}
\Omega=-[1-e^{2L/l_{sf}^{N}}\prod^{2}_{i=1}(2\rho_{i}^{I}+1)]\times(P^{I}_{2}\eta^{I}_{2}+\alpha^{F}_{2}\eta^{F}_{2}) \notag \\
\times[1-(2\rho_{2}^{I}+1)(2\eta_{1}^{I}+2\eta_{1}^{F}+1)e^{2L/l_{sf}^{N}}] \notag \\
\end{eqnarray}

As we can see, for $\delta (\theta )$, the angular dependence comes
only from the term $R(\theta)/cos\theta$. As the numerator of
$R(\theta)$ also has a term of $cos\theta$, the angular dependence
of $\delta (\theta )$ is only determined by $sin^{2}\theta$.
Obviously, $\delta(0^{o})$ exactly equals to $\delta(180^{o})$,
which is quite different from that in conventional FM spin valve.
Such symmetry comes from the fact that no electric current flows in
FM2 lead. Detailed analysis will be given in the next part.
\section{Numerical Results and Discussion}

\begin{figure}[tbp]
\includegraphics[width=8cm]{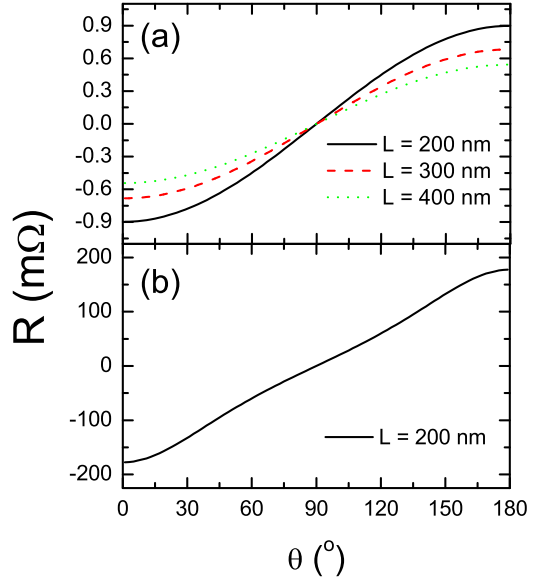}
\caption{(color online) Angular dependence of AMR in NLSV, (a) for
metallic contact with different space $L$, (b) for tunnelling
contact with $L=200nm$.} \label{fig:2}
\end{figure}

The AMR of NLSV could be directly measured in experiments and be used to
test our theoretical prediction. For the non-collinear NLSV we considered,
the permalloy is taken as the ferromagnetic leads while the cooper as normal
lead. The material parameters entering our formulism adapt the values
extracted from the experiments.\cite{Py_Cu_parameter,J_Bass_3M_CPP_GMR} The
parameter $G_{\uparrow \downarrow }^{I}$ for the Py/Cu contact follows that
in Ref.\onlinecite{Universial_AMR}. For tunnelling contact, according to
\emph{ab.initio} calculation\cite{torque_Ke} the contact resistance could be
taken 11 times that of metallic contact for thick barrier and the mixing
conductance $G_{\uparrow \downarrow }^{I}$ is almost unchanged. The contact
area of FM2$/$NM is assumed to be constant with variation of $\theta $. The
two ferromagnetic leads are also assumed to be identical.

\textbf{\emph{AMR in NLSV}}: The AMR with different distance $L$ between FM1
and FM2 is shown in Fig.\ref{fig:2}(a) for metallic contact. It is found
that the absolute value of $R(\theta )$ decreases with increase of $L$. This
is due to the fact that the spin-flip scattering could kill the spin memory
in normal metal. With increasing $L$, the spin accumulation at the NM side
of FM2$/$NM contact decreases. For metallic system, the dimensionless
parameters $\eta ^{I}$, $\eta ^{F}$, and $\rho ^{I}$ are always less than
unit. Therefore, the third terms in denominator in Eq.(\ref{eq-6}) can be
neglected compared with other two terms. As the consequence, AMR is govern
by the nominator of Eq.(\ref{eq-6}), and gives a cosine line shape of $%
R(\theta )$, which was discussed by Levy et al.,\cite{levy}
recently.
\begin{figure}[tbp]
\includegraphics[width=7.5cm]{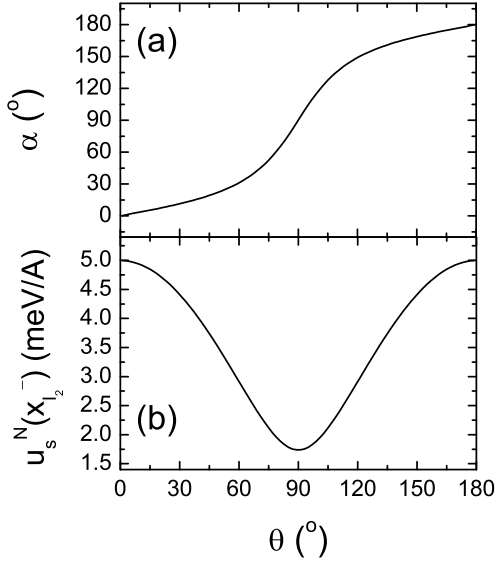}
\caption{(a)Angular dependence of the relative angle $\protect\alpha $
between the direction of $\mathbf{u}_{s}^{N}(x_{I_{2}}^{-})$ and $\mathbf{m}%
_{2}$. (b)Angular dependence of the magnitude of $\mathbf{u}%
_{s}^{N}(x_{I_{2}}^{-})$ normalized by the injected current $I_{o}$.}
\label{fig:3}
\end{figure}

Fig.\ref{fig:2}(b) presents the AMR with tunnelling contact, where $L=200nm$%
. Because the spin transport is dominated by contact, the line shape
shows a very difference from that with a metallic contact. Such
variation of line shape of AMR implies that $R(\theta)$ decreases
more quickly in tunnelling contact when FM1 and FM2 in noncollinear
configuration. The reason is that the noncollinear spin accumulation
in NM resistor could leak out more efficiently with tunnelling
contact comparing with metallic contact.  It is known that the drift
of the noncollinear spin accumulation across the FM/NM contact is
dominated by
$G_{\uparrow\downarrow}^{I}/G^{I}$.\cite{Brataas_circuit} For
metallic contact, $G^{I}$ is comparable to
$G_{\uparrow\downarrow}^{I}$. For tunnelling contact, in spite that
$G^{I}$ is very small, $G_{\uparrow\downarrow}^{I}$ still has
similar magnitude to that in metallic contact.\cite{torque_Ke} Then,
due to the fast leak of the noncollinear spin accumulation with
tunnelling contact, $R(\theta)$ decays more quickly when deviating
from collinear configuration. In this case, AMR in NLSV could be
more sensitive to the quantity $\rho ^{I}$. This makes an effective
way to extract the mixing conductance $G_{\uparrow \downarrow }^{I}$
from experiment with the tunnelling contacts.

For both the metallic and the tunnelling contacts, we have
$R(90^{o})=0$ shown in Fig.\ref{fig:2}, which is due to the
vanishing voltage difference across FM2$/$NM contact when $\theta
=90^{o}$. This does not mean that the spin accumulation at the NM
side of FM2$/$NM is vanishing. After the injection from FM1, the
electrons will be polarized along $\mathbf{m}_{1}$ at first. For
$\theta =90^{o}$, the spin of electrons arriving at the NM side of
FM2$/$NM contact is perpendicular to the magnetization of FM2. The
induced voltage across FM2$/$NM will not change when the
magnetization of FM2 reversed.

\begin{figure}[tbp]
\includegraphics[width=7cm]{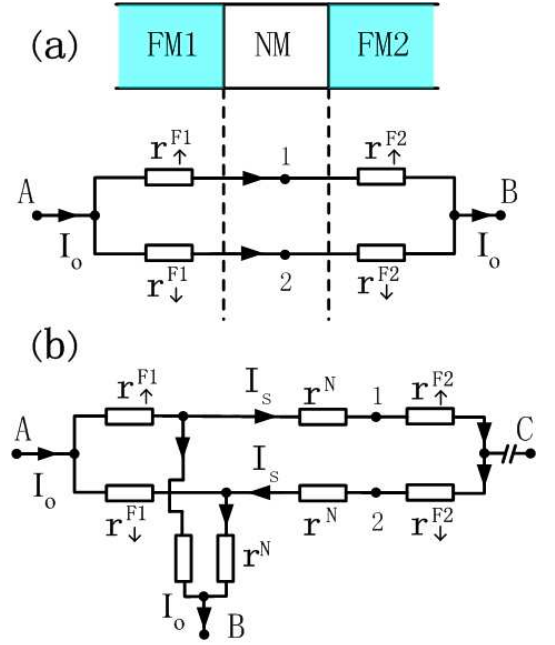}
\caption{(color online) (a)Side view of local spin valve and equivalent
circuit of collinear magnetic configuration, (b) equivalent circuit of
collinear magnetic configuration for NLSV, where A,B,C correspond to the
same points in Fig.\protect\ref{fig:1}. }
\label{fig:5}
\end{figure}

\textbf{\emph{Spin Accumulation in NLSV}}: As the Im$G_{\uparrow \downarrow
}^{I}$ related term in Eq.(\ref{eq-10}) is disregarded in metallic system,
the spin accumulation $\mathbf{u}_{s}^{N}(x_{I_{2}}^{-})$ at the NM side of
FM2$/$NM is in the plane spanned by $\mathbf{m}_{1}$ and $\mathbf{m}_{2}$.
Fig.\ref{fig:3}(a) presents the $\theta $ dependence of the relative angle $%
\alpha $ between direction of $\mathbf{u}_{s}^{N}(x_{I_{2}}^{-})$ and $%
\mathbf{m}_{2}$. As we discussed above, when $\theta =90^{o}$, FM2$/$NM is
equivalent to an unpolarized contact and $\alpha =90^{o}$ is expected. Fig.%
\ref{fig:3}(b) gives the $\theta $ dependence of the magnitude of $\mathbf{u}%
_{s}^{N}(x_{I_{2}}^{-})$ normalized by the injected current
$I_{o}$. The magnitude reaches its minima at $\theta =90^{o}$
while gives identical value for parallel($\theta =0^{o}$) and
antiparallel($\theta =180^{o}$) configurations, which is quite
different from that in LSV. Such discrepancy can be identified
through the equivalent circuit of LSV and NLSV as shown in
Fig.\ref{fig:5}, where the circuits follow the collinear magnetic
configuration of LSV and NLSV with $r_{\uparrow (\downarrow
)}^{F}=l_{sf}^{F}/(\sigma _{\uparrow (\downarrow
)}^{F}S^{F})+1/G_{\uparrow (\downarrow )}^{I}$ and
$r^{N}=2l_{sf}^{N}/(\sigma ^{N}S^{N})$.

For both types of spin valve, the spin accumulation in the normal metal
equals to the potential difference between node 1 and node 2 (see Fig.\ref%
{fig:5}). In LSV, as the particle current flows from FM1 to FM2, the
switching of magnetization of FM2 will interchange the resistors $%
r_{\uparrow }^{F2}$ and $r_{\downarrow }^{F2}$, which could change the
potential on node 1 and node 2. However, in NLSV, the electrical current $%
I_{o}$ flows from electrode $A$ to electrode $B$ and no net electrical
current flows to the detection lead FM2, namely, electrode $C$. Only spin
current, which is denoted as $I_{s}$, flows to F2. It is obvious that the
interchange of $r_{\uparrow }^{F2}$ and $r_{\downarrow }^{F2}$ in Fig.\ref%
{fig:5}(b) do not affect the current $I_{s}$. As a result, the potential
difference between nodes 1 and 2 will not be changed. In non-collinear
magnetic configuration of NLSV, the equivalent circuit in Fig.\ref{fig:5}(b)
is not valid anymore. Due to non-collinear transport, more channels will be
opened\cite{Brataas_circuit} and new resistors could directly connect the
node 1 and node 2. The potential difference will be changed with variation
of the direction of FM2.

\begin{figure}[tbp]
\includegraphics[width=8cm]{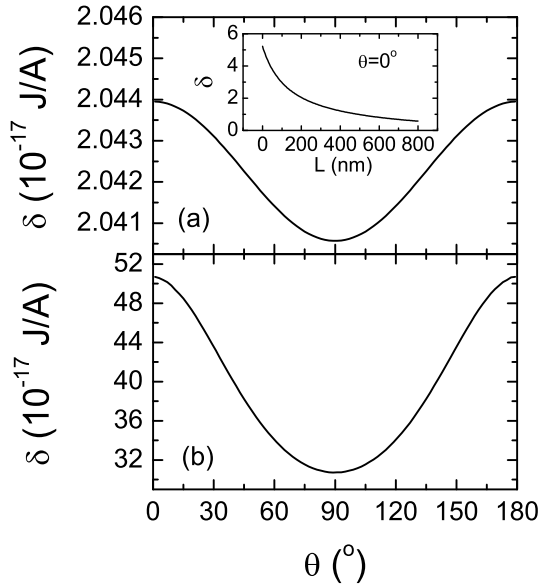}
\caption{Angular dependence of spin torques on FM2 of NLSV with
$L$=200nm, (a) for the metallic contact and (b) for the tunnelling
contact. Inset of (a) gives the space $L$ dependence of $\delta
(0^{o})$.} \label{fig:4}
\end{figure}

\textbf{\emph{STT in NLSV}}: The spin accumulation near the
FM2$/$NM contact could induce a STT on the detection lead FM2. For
the metallic and tunnelling contacts, we have calculated $\delta
(\theta )$ and presented the results in Fig.\ref{fig:4}. Even
though $\tau $ is always zero when two magnetization are aligned
collinearly, $\delta (0^{o})$
and $\delta (180^{o})$ show nonzero values as in LSV. For typical space of $%
L=200nm$, the STT obtained is smaller than that in
LSV,\cite{Manschot_STT} but still in the same order of magnitude.
The spin-flip scattering in NM could suppress STT as the space $L$
increases. The space $L$ dependence of $\delta (0^{o})$ is shown
in the inset of Fig.\ref{fig:4}(a). As we can see, sizable STT
could be expected even in NLSV with $L$ comparable or less than
$l_{sf}^{N}$ which is 700nm in this study. The $\delta $ in the
NLSV with tunnelling contacts could be even larger because in this
case the contact dominates the spin transport and with higher spin
injection efficiency\cite{Jedema_nature_2002} the spin
accumulation in the normal electrode is essentially enhanced per
unit current.

Interestingly, the spin torques in NLSV still change their signs
when the injected current is reversed. As the electrical
(electron) current $I_{o}$ injected from electrode A to B as shown
in Fig.\ref{fig:5}(b), for the materials we discussed
($r_{\uparrow }^{F}<r_{\downarrow }^{F}$), spin accumulation
parallel to $\mathbf{m}_{1}$ could be built up in NM, which will
exert STT to FM2. When we reverse the current $I_{o}$, electrons
come from electrode B to A and the spin dependent reflection at
FM1$/$NM contact will built up a spin accumulation antiparallel to
$\mathbf{m}_{1}$ in NM. So the STT changes it sign.

Contrasting to the STT in LSV,\cite{Boltzmann_torque} $\delta (\theta )$ is
symmetrical for the parallel and antiparallel magnetic configuration of FM1
and FM2. See Eq.(\ref{eq-8}), the symmetry comes from the symmetrical
angular dependence of $\mathbf{u}_{s}^{N}(x_{I_{2}}^{-})$ shown in Fig.\ref%
{fig:3}(b). This implies that the critical current should be identical for
parallel to antiparallel and antiparallel to parallel.

Switching behavior in the NLSV has been observed by Kimura et al.,\cite%
{Kimura_NLSV_switch} even though they only observed antiparallel to parallel
switching, where the NM lead in Fig.\ref{fig:1} is replaced by a NM cross
and the FM leads are placed on two opposite arms of the cross. The spin
accumulation could leak from those arms not in contact with ferromagnetic
leads. Therefore, the magnitude of STT could be 2-3 times weaker than that
in NLSV discussed here.

\section{Summary}

Based on the diffusion equation and magnetoelectronic circuit
theory, the non-collinear spin transport in NLSV is treated
analytically and numerically in the diffusive regime. The
analytical expression of AMR defined in NLSV is derived. For the
system with metallic contacts, the AMR gives a cosine function
like angular dependence. For the system with tunnelling contacts,
the AMR shows complicate angular dependence and could be used to
extract mixing conductance from experiment. The STT in NLSV has
the same order of magnitude as that in LSV but shows qualitative
difference in the angular dependence. The STT in NLSV is found to
be symmetrical when the two FM leads parallel and antiparallel to
each other. The symmetry comes from the fact only spin current
flows across the detection lead. Our study implies that the
critical current of magnetization switching in NLSV could be
identical for parallel configuration and antiparallel
configuration.

\begin{acknowledgments}
Y.X. and K.X. acknowledge financial support from NSF(10634070) and
MOST(2006CB933000, 2006AA03Z402) of China; Z.S.M. acknowledge
financial support from NNSFC Grant No. 10674004 and NBRP-China
Grant No. 2006CB921803.
\end{acknowledgments}

\end{document}